\documentstyle[12pt,aasms4,flushrt]{article}

\newcommand{\Lya}{\mbox{Ly$\alpha$} }

\def\gsim{\;\rlap{\lower 2.5pt
 \hbox{$\sim$}}\raise 1.5pt\hbox{$>$}\;}
\def\lsim{\;\rlap{\lower 2.5pt
   \hbox{$\sim$}}\raise 1.5pt\hbox{$<$}\;}
\newcommand\beq{\begin{equation}}
\newcommand\eeq{\end{equation}}

\def\v{\vspace{-0.1in}}

\begin{document}
\Large 
\centerline{\bf Spectral signature of cosmological infall}
\centerline{\bf of gas around the first quasars} 
\normalsize 
\author{\bf
Rennan Barkana$^{\star}$ and Abraham Loeb$^{\dagger}$}

\noindent
$^{\star}$ School of Physics and Astronomy, Tel Aviv University,
Tel Aviv 69978, ISRAEL

\noindent
$^{\dagger}$ Institute for Advanced Study, Princeton, NJ 08540, USA;
on sabbatical leave from the Astronomy Department, Harvard University,
Cambridge, MA 02138, USA

\vskip 0.2in 
\hrule 
\vskip 0.2in 
{\bf Recent observations have shown that, only a billion years after
the Big Bang, the Universe was already lit up by bright
quasars\cite{f1} fuelled by the infall of gas onto supermassive black
holes. The masses of these early black holes are inferred from their
luminosities to be $> 10^9$ solar masses ($M_{\odot}$), which is a
difficult theoretical challenge to explain. Like nearby quasars, the
early objects could have formed in the central cores of massive host
galaxies. The formation of these hosts could be explained if, like
local large galaxies, they were assembled gravitationally inside
massive ($> 10^{12} M_{\odot}$) halos of dark matter\cite{BL1}. There
has hitherto been no observational evidence for the presence of these
massive hosts or their surrounding halos. Here we show that the cosmic
gas surrounding each halo must respond to its strong gravitational
pull, where absorption by the infalling hydrogen produces a distinct
spectral signature. That signature can be seen in recent
data\cite{SDSSz5}$^,$\cite{z6p3}.}

We model the effect of resonant \Lya absorption by infalling gas on
the quasar light (Figure~1). For each quasar we consider the history
of the formation of its host halo from an initial positive overdensity
of dark matter. For the initial surrounding density profile at high
redshift we adopt the typical profile expected around the dense region
that collapses to form the halo\cite{le95}. This profile as well as
the approximation of a spherical geometry are particularly
satisfactory for the very rare and massive halos under
consideration. We calculate gas infall down to the radius of the
accretion shock, and neglect any
\Lya absorption due to the post-shock gas. The hot ($\ga 10^7$K)
post-shock gas should be fully-ionized by collisions.  Part of it is
expected to subsequently cool and collapse onto the galactic disk, but
Compton heating by the quasar should keep the virialized gas hotter
than $\sim 10^6$K. Even if a thin cold shell of shocked gas remains,
it will not change the basic pattern produced by infalling gas since
the post-shock gas no longer has a high infall velocity.

In order to predict the \Lya absorption around a quasar we must
estimate the mass of its host halo. A tight correlation has been
measured in local galaxies between the mass of the central black hole
and the bulge velocity
dispersion\cite{BHlocal1}$^,$\cite{BHlocal2}. This relation also fits
all existing data on the luminosity function of high-redshift quasars
within a simple model\cite{Wyithe} in which quasar emission is assumed
to be triggered by mergers during hierarchical galaxy formation. We
use the best-fit\cite{BHlocal1}$^,$\cite{BHlocal2}$^,$\cite{Wyithe}
relation, in which the black hole mass in units of $10^8 M_{\odot}$
($M_8$) is related to the circular velocity at the halo boundary in
units of 300 km s$^{-1}$ ($V_{300}$) by $M_8 = 1.5 (V_{300})^5$. For
the typical quasar continuum spectrum, we adopt a power-law shape of
$F_{\nu} \propto \nu^{-0.44}$ in the rest-frame range 1190--5000\AA\
based on the Sloan Digital Sky Survey (SDSS) composite
spectrum\cite{SDSScomp}, and $F_{\nu} \propto \nu^{-1.57}$ at
500--1190\AA\ using the composite quasar spectrum from the {\it Hubble
Space Telescope}\cite{EUVcomp}. Based on observations in soft
X-rays\cite{Xcomp}, we extend this power-law towards short
wavelengths.

We assume that the brightest quasars shine at their Eddington
luminosity, and we note that for the SDSS composite
spectrum\cite{SDSScomp}, the total luminosity above 1190\AA\ equals
1.6 times the total continuum luminosity at 1190--5000\AA. Thus,
ionizing photons stream out of the quasar's galactic host at the rate
$\dot{N} = 1.04\times 10^{56} M_8$ s$^{-1}$. We infer the black hole
mass using the observed continuum at 1350\AA, with the conversion
$F_{\nu} = 1.74 \times 10^{30} M_8$ erg s$^{-1}$ Hz$^{-1}$. Given that
helium is doubly-ionized by the quasar, the frequency-averaged
photoionization cross-section of hydrogen is for our template spectrum
$\bar{\sigma}_H = 2.3 \times 10^{-18}$ cm$^2$. The double ionization
of helium increases the recombination rate due to the extra electrons,
and also produces a characteristic gas temperature of $\sim 1.5\times
10^4$ K in regions that had already been reionized by a softer
ionizing background\cite{HeRadTr}.

Only a limited number of published spectra are currently available for
a clear test of our predictions. First, only the brightest quasars,
which reside in the most massive halos, produce strong infall over a
large surrounding region. The infalling gas density scales as
$(1+z)^3$ and the shock radius scales as $(1+z)^{-1}$, so that the
absorption optical depth (see below) scales roughly as $(1+z)^4$ and
the absorption feature is much weaker at low redshift. Even at high
redshifts, measuring the detailed \Lya line profile is possible only
in a high resolution spectrum with an extremely high signal-to-noise
ratio.

Figure~2 shows two particularly high-quality spectra and compares them
with our model predictions. Both spectra show our predicted
double-peak pattern and disagree with the single-peaked profile
predicted by previous models that ignored infall. In particular,
models that assume a pure Hubble flow firmly predict no absorption at
all at positive velocities and only a gradual decline in the
transmitted flux toward negative velocities. This slow decline is due
to the fact that in these models the gas closest to the quasar
produces no absorption since it is fully ionized by the strong
ionizing intensity of the quasar. More distant gas recedes along with
the universal expansion and absorbs only at negative velocities. Our
model, in contrast, firmly predicts a sharp flux cutoff located at a
positive velocity. This flux drop corresponds to absorption by gas
just outside the accretion shock, and the positive velocity of the
cutoff corresponds to the infall velocity. The strong absorption
caused by this gas despite exposure to the ionizing flux from the
quasar is due to the high density of the infalling gas. This gas is
expected to be $\sim 10$--30 times denser than the cosmic mean after
having fallen toward the quasar host halo over the history of its
formation.

The infall velocity is proportional to the halo circular velocity and
thus\cite{BL1} to $M^{1/3} (1+z)^{3/2}$, in terms of the halo mass $M$
and the redshift $z$. The actual physical distance of the accretion
shock in the model is 86 kpc and 80 kpc for the quasars at $z=4.795$
and $z=6.28$, respectively. Our model with infall also fits
approximately the secondary peak that is observed on the blue side of
the main peak (i.e., at lower velocity). This blue peak corresponds
closely to the point of weakest absorption by the infalling gas,
although the profile of the intrinsic quasar emission also affects
somewhat the precise location of this peak. In this region we do not
expect to fit the flux profile in full detail; our model averages over
all lines of sight and possible quasar positions, but observable
random fluctuations around our predicted mean are expected in each
specific spectrum.

However, once we average over the density fluctuations, the prediction
of a second peak rather than a smooth fall-off is generic and
insensitive to the detailed model assumptions. At a distance $R$ from
the quasar the resonant optical depth depends on $\rho^2 R^2$, where
$\rho$ is the density including infall; one factor of $\rho$ comes
from the total gas density, the second comes from the \ion{H}{1}
fraction which increases with $\rho$ due to recombinations, and the
$R$-dependence results from the $R^{-2}$ decline of the ionizing
intensity of the quasar. Thus, a second peak should appear as long as
infall produces a density profile falling off faster than $1/R$ (our
model predicts $R^{-3/2}$) before asymptoting to the cosmic mean value
of unity.  We note that in our model the position of the blue peak
corresponds to absorption by gas at $R=0.45$ Mpc ($z=4.795$) or 0.41
Mpc ($z=6.28$).

Even in each particular spectrum, the more distant region where the
flux drops toward zero can be modelled much more robustly. In
particular, significant flux is observed at $\Delta V = -2000$ km
s$^{-1}$ ($z=4.795$) and $\Delta V = -3000$ km s$^{-1}$ ($z=6.28$),
respectively. These relatively distant regions are only weakly
affected by infall and the observed positions translate to a distance
from the quasar of 3.8 Mpc ($z=4.795$) or 4.2 Mpc ($z=6.28$). In
models that do not include a clump distribution, the optical depth at
these positions is $\ga 3$, which means that the observed flux
requires an intrinsic unabsorbed flux that is 20 times greater. This
is clearly impossible regardless of any uncertainties about the
intrinsic line shape and the quasar continuum level. To explain the
observed flux, the ionizing intensity of the quasar as determined by
our template spectrum from the observed continuum would have to be too
low by a factor $> 2$. However, our full model accounts for the fact
that part of the cosmic gas falls into dense sheets and filaments and
leaves the rest with a density below the cosmic mean. The resonant
\Lya absorption is made up of the separate contributions of gas
elements at a variety of overdensities, and since gas in low-density
regions absorbs very weakly, clumping actually increases the mean
transmission. Thus, our model naturally accounts for the flux observed
far from the quasar, with no change required in the quasar spectrum.

Our conclusions are insensitive to the question of whether the
\ion{H}{2} region of the highest redshift quasar is surrounded by a
region of neutral hydrogen (due to the fact that the universe had not
been fully reionized by $z=6.28$\cite{z6p3b}) or not\cite{me}; a
distant neutral region would only add on the IGM damping
wing\cite{jordi98} which produces a smooth, gradual suppression that
should not alter the basic double-peak pattern. We note that the
quasar may possess a velocity offset relative to Hubble flow due, for
example, to a violent galactic merger that had originally activated
the quasar. However, the close fit that we find between the predicted
accretion shock position and the observed flux drop is evidence
against the presence of a large velocity offset in the two quasars we
have considered. We note as well that if the observed absorption
pattern were due to dense gas clouds within the galaxy (for example,
in the region feeding the central black hole), then the absorbing gas
would be expected to contain heavy elements such as carbon, oxygen,
and nitrogen. This gas would then be expected to absorb other emission
lines of the quasar in addition to
\Lya. Such associated absorption is absent in the $z=4.795$ quasar
which has several emission lines with well measured
profiles\cite{SDSSz5}, suggesting that the absorbing gas has a
near-primordial composition as expected for intergalactic gas.

Our models provide the first direct evidence that two characteristic
properties of quasars at low redshift are also applicable to bright
quasars in the early universe. These properties include the quasar
spectral template, which determines the ionizing intensity of the
quasar, and the relation between black hole mass and halo velocity
dispersion, which we have used to determine the host halo mass. Both
observed spectra show a blue peak of about $75\%$ of the height of the
red (positive velocity) peak, and this is roughly matched by the
models. However, if we were to increase the ionizing intensity by an
order of magnitude then we would predict a blue peak at least of equal
height to the other peak. If, instead, we decreased the ionizing
intensity by an order of magnitude then the resulting blue peak would
be under $50\%$ of the height of the red peak and the transmitted flux
would decrease to zero toward negative velocities much faster than is
observed. Similarly, if we varied the assumed halo mass by more than
an order of magnitude then the resulting absorption profile in each
quasar would disagree with the data. High-redshift quasars could in
principle be much fainter intrinsically than they appear, if they are
magnified by gravitational lensing\cite{Wyithe2}; our limits on the
ionizing intensity, however, suggest that the two quasars we have
modelled cannot be magnified by a factor $\ga 10$.

We can also estimate from the data the total gas infall rates into
these massive galaxies. The positions of the accretion shocks imply in
our models infall velocities of 400--550 km s$^{-1}$ and shock radii
of 80--90 kpc. Since gas at this radius is expected to have a density
of $\sim 20$ times the cosmic mean density\cite{le95}, we obtain
accretion rates of $1300\, M_{\odot}$ yr$^{-1}$ ($z=4.795$) and
$2900\, M_{\odot}$ yr$^{-1}$ ($z=6.28$), respectively. At these rates,
the host galaxies of these two quasars could have been assembled in
2--3$\times 10^8$ yr, consistent with the $9 \times 10^8$ yr age of
the universe at $z=6.28$. Future comparison of our model to the {\it
average}\, \Lya absorption profile of a statistical sample of bright,
early quasars with similar luminosities and redshifts should allow us
to fit the details of the absorption spectrum and refine our
quantitative conclusions.

\vskip 0.2in
\noindent
{\bf Acknowledgments} The authors thank Ed Turner and Hagai Netzer for
helpful discussions, and are grateful for the hospitality of the
Institute for Advanced Study where this work was completed. RB
acknowledges the support of an Alon Fellowship at Tel Aviv University
and of the Israel Science Foundation.  AL acknowledges support from
the Institute for Advanced Study and a John Simon Guggenheim Memorial
Fellowship. This work was also supported by the National Science
Foundation.

\vskip 0.2in
\noindent 
{\bf Correspondence} and requests for materials should be addressed to
R.B. (email: barkana@wise.tau.ac.il) or A.L. (email: loeb@ias.edu).

\vskip 1in
 
\begin{figure*}[hptb]
\plotone{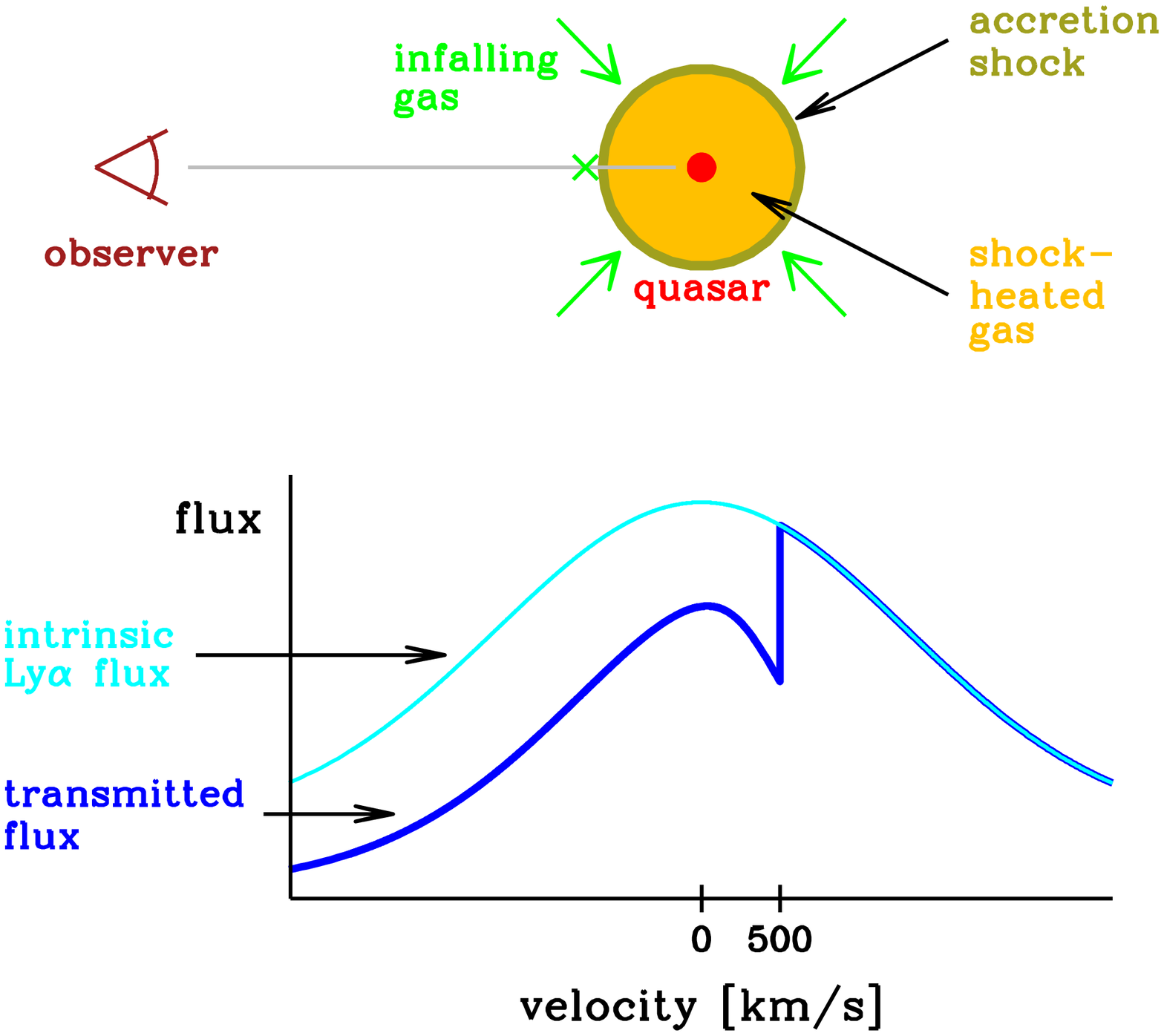}
\end{figure*}

\begin{figure*}[hptb]
\caption{Schematic illustration of how infall produces a unique
spectral signature. A quasar forms inside a galaxy that lies at the
center of a massive dark matter halo. A large volume of gas responds
to the strong gravitational pull and falls toward the massive halo. As
infalling gas impacts on the galactic gas, a strong accretion shock
forms. The intrinsic quasar \Lya emission is partially absorbed by the
infalling pre-shock gas. In particular, a sharp flux drop is caused by
gas (marked with an $\times$) that is about to hit the accretion
shock. This gas is falling toward the quasar at 500 km s$^{-1}$ in
this sketch. We calculate spherically-symmetric infall\cite{gg72} and
set the accretion shock radius to 1.15 times the halo
boundary\cite{2ndInfall}, although our results are not altered
substantially as long as the shock radius is close to the halo virial
radius. Three-dimensional hydrodynamic simulations show that the most
massive halos at any time in the universe are indeed surrounded by
strong, quasi-spherical accretion
shocks\cite{EliSPH}$^,$\cite{Abel}. Preliminary evidence has been
found for such shocks around nearby galaxy
clusters\cite{sm02}$^,$\cite{lw00}. We model resonant \Lya absorption
by intergalactic hydrogen\cite{GP} that is partially ionized by the
radiation produced by the quasar\cite{prox2}. In addition to the
overall infall pattern\cite{le95} we include a realistic distribution
of gas density fluctuations\cite{zoltan}. For the distribution of gas
clumps we adopt an analytical fit to numerical
simulations\cite{clumping}. We assume that the clumps are optically
thin, and find the neutral fraction separately for each clumping
density based on ionization equilibrium with the quasar ionizing
flux. We then calculate the mean \Lya transmission averaged over the
clump distribution. Note that the absorbed \Lya photons are re-emitted
from a large \Lya halo\cite{LyaHalo1}$^,$\cite{LyaHalo2} that is too
faint to significantly affect current observations.}
\end{figure*}

\begin{figure*}[hptb]
\plotone{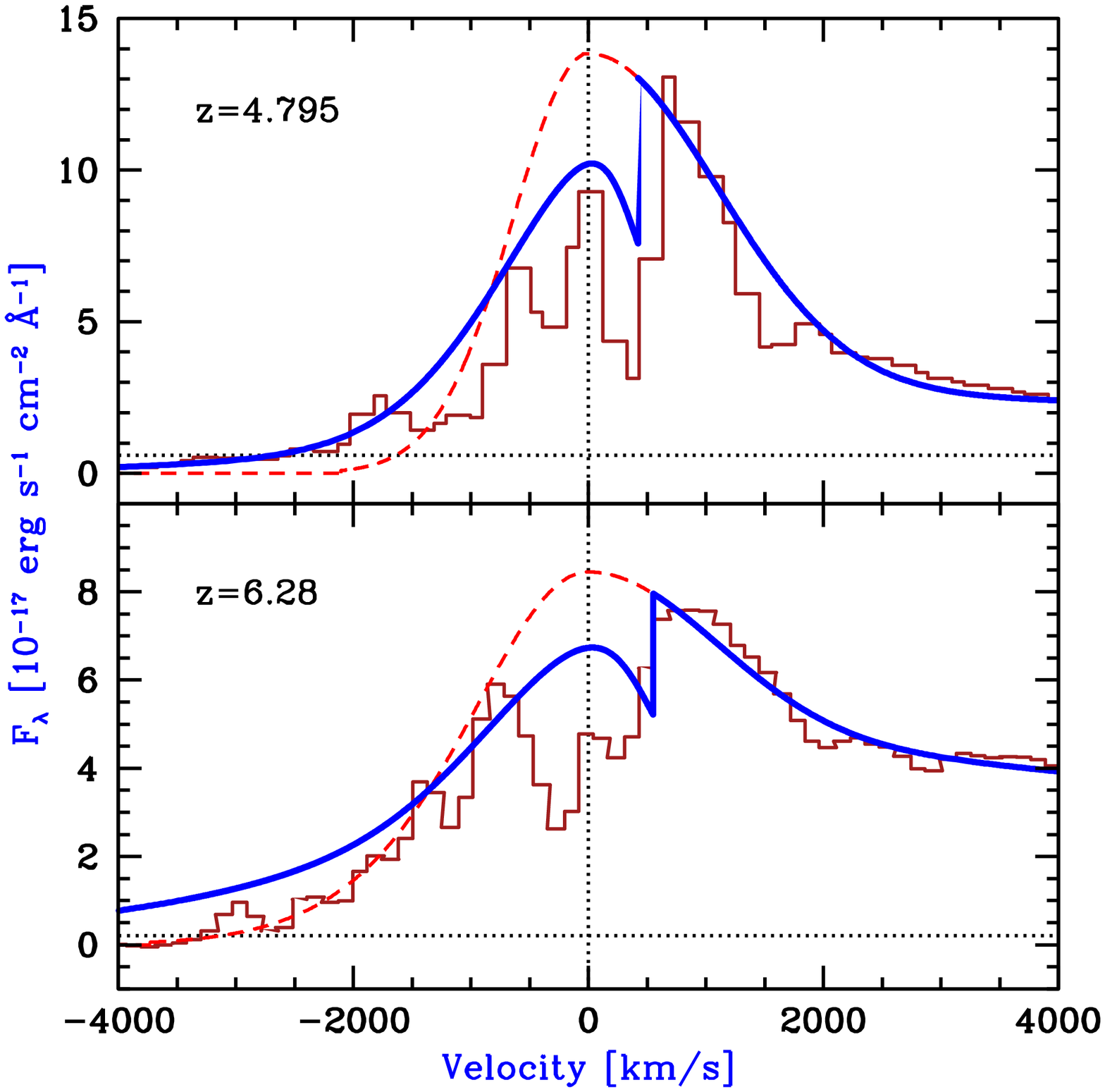}
\end{figure*}

\begin{figure*}[hptb]
\caption{Comparison between models of cosmological infall and observed
quasar spectra. The upper panel considers the redshift $4.795\pm
0.004$ quasar SDSS 1122-0229\cite{SDSSz5}; based on the observed
redshift and continuum level our model implies a $4.6 \times 10^8
M_{\odot}$ black hole residing in a $2.5 \times 10^{12} M_{\odot}$
host halo.  The lower panel considers the redshift $6.28\pm 0.02$
quasar SDSS 1030+0524\cite{z6p3}, for which our model implies a $1.9
\times 10^9 M_{\odot}$ black hole residing in a $4.0 \times 10^{12}
M_{\odot}$ host halo (we do not use a second spectral observation of
this same source\cite{px} since it appears to have a significantly
lower signal-to-noise ratio). In each panel, the histogram shows the
observed spectrum, the dashed line shows previous models that assume a
uniform expanding universe, and the solid line shows our model which
includes cosmological infall as well as a realistic distribution of
gas clumps. Note that the velocity is measured relative to the quasar,
where negative velocity means motion towards us. In each panel, the
vertical dotted line shows the position of the \Lya wavelength at the
source redshift, and the horizontal dotted line shows the flux level
of the highest transmission peaks seen in parts of the spectrum
corresponding to the average intergalactic medium (i.e., at velocities
more negative than -4000 km s$^{-1}$). We assume an intrinsic emission
line given by a sum of two Gaussian components, a form which best fits
the line shape of most quasars at low redshift\cite{Hagai2}; we fix
the parameters for each quasar based on the unabsorbed part of the
\Lya line at velocities above 500 km s$^{-1}$. With this approach the
models do not include any free parameters. Particular quasars are
expected to show fluctuations around our predicted absorption profile,
since our model averages over random lines of sight and density
fluctuations. Throughout this paper we assume the standard
cosmological parameters $\Omega_{m}=0.3$, $\Omega_{\Lambda}=0.7$,
$\Omega_b=0.05$, $H_0=70$ km s$^{-1}$ Mpc$^{-1}$, and $n=1$.}
\end{figure*}

\end{document}